\documentclass[12pt,twoside]{article}
\usepackage{a4,amsmath,latexsym,exscale,russian,amssymb}
\input epsf
\newcommand {\al}   {\alpha}       \newcommand {\bt}  {\beta}
\newcommand {\g }   {\gamma}       \newcommand {\G }  {\Gamma}
       
\newcommand {\z }   {\zeta}

          \newcommand {\x }  {\xi}
\newcommand {\s }   {\sigma}      %\newcommand {\r }   {\rho}
         
\newcommand {\f }   {\varphi}      
         
\newcommand {\Lm}   {\Lambda}      
       
\newcommand {\pl}   {\partial}     
\newcommand   {\const}{{\sf const}}   
       \renewcommand {\det}{{\sf det}}

\renewcommand {\ln}{{\sf ln}}         
\renewcommand {\exp}{{\sf exp}}     % \renewcommand {\log}{{\sf log}}
       \renewcommand {\cos}{{\sf cos}}
\newcommand   {\tg}{{\sf tg}}         \newcommand   {\ctg}{{\sf ctg}}

\newcommand   {\sh}{{\sf sh}}         \newcommand   {\ch}{{\sf ch}}
\renewcommand {\tanh}{{\sf th}}       \newcommand   {\cth}{{\sf cth}}
   
\newcommand   {\arcth}{{\sf arcth}}  %\newcommand   {\cth}{{\sf cth}}

   \newcommand {\MR}  {{\mathbb R}}

\newcommand {\CC }  {{\cal C}}
\newtheorem{Theorem}  {Theorem}
\begin{document}
\title     {Global solutions in gravity.\\  Lorentzian signature.}
\author    {M. O. Katanaev
            \thanks{E-mail: katanaev@mi.ras.ru}\\ \\
            \sl Steklov Mathematical Institute,\\
            \sl Gubkin St. 8, 117966, Moscow, Russia}
\date      {}
\maketitle
\begin{abstract}
  The constructive method of conformal blocks is developed for the
  const\-ruc\-tion
  of global solutions for two-dimensional metrics having one Killing
  vector. The method is proved to yeild a smooth universal covering space
  with a smooth pseudo-Riemannian metric. The Schwarzschild,
  Reisner--Nordstrom solutions, extremal black hole, dilaton black hole,
  and constant curvature surfaces are considered as examples.
\end{abstract}
%********************************************************************
\section{Introduction}
%********************************************************************
A space-time in gravity models is a differentiable manifold which has
to satisfy at least two requirements. Firstly, the Lorentz signature
metric satisfying some system of equations of motion is to be given on it.
Secondly, the manifold has to be maximally extended along extremals.
The last requirement means that any extremal can be either continued to
infinite value of the canonical parameter in both directions or at a finite
value of the canonical parameter it ends up at a singular point where
one of the geometric invariants, for example, the scalar curvature
becomes infinite. Space-time with a given metric satisfying both
requirements is called a global solution in gravity. This definition is
invariant and do not depend on the coordinate system because the
canonical parameter is invariant and defined up to a linear transformation.
The Kruskal--Szekeres extention \cite{Kruska60,Szeker60} of the
Schwarzschild solution \cite{Schwar16} is a well
known example of nontrivial global solution in general relativity.

The necessity to construct global solutions, that is, manifolds themselves
and metrics given on them, is related to the invariance of gravity models
under general coordinates transformations. As a rule some coordinates are
fixed when solving the equations of motion, and this reduces the number of
unknown functions. On the other hand fixing of the coordinate system
means that excluding trivial examples the obtained solution is a local
one and represents only a part of some larger space-time. Note also that
only the global structure of a manifold allows one to give the physical
interpretation of a solution to the equations of motion for the metric
because it is invariant.

The problem setting by itself differs from standard problems in
mathematical physics where a manifold is assumed to be given in advance,
and the task is to solve one or the other boundary value problem.
In contrast to this the boundary conditions are replaced by the
requirement of completeness, and the manifold by itself is
constructed along with the equations solving. It is interesting
to analyse the relation between the theory of global solutions
and general theory of dynamical systems, the basics of which were
founded by N.~N.~Bogolubov \cite{Bogolu39}.

Construction of global solutions in gravity is a difficult problem
because besides a solution of the equations of motion which is a
difficult task by itself includes solution of the equations for extremals
analisys of their completeness, and extention of the manifold in the
case of its incompleteness. In general relativity only a small number
of global solutions is known in account of the complicated equations of
motion (see review \cite{Carter73}). In two-dimensional gravity models
attracting much interest last years the situation is simpler, and all
global solutions were succeeded to be found \cite{Katana93A} in
two-dimensional gravity with torsion \cite{KatVol86}, and in a large
class of dilaton gravity models too \cite{KloStr96C,KaKuLi96,KaKuLi97}.
Constractive method of conformal blocks was proposed \cite{Katana93A}
for two-dimensional gravity with torsion in the conformal gauge, the
${\cal C}^2$ continuity being proved on horizons along which local
solutions were sewed together. The proof was based only on the analisys of
asymptotics near horizons, and therefore it is applicable to other models.
The use of Eddington--Finkelstein coordinates in the analisys of global
solutions for a wide class of two-dimensional models including dilaton
gravity \cite{KloStr96C} allowed to prove the smoothness of global solution.
Analisys of the equations of motion in general relativity showed that
solutions having the form of a warped product of two surfaces can also be
explicitly constructed and classified \cite{KaKlKu99}. In fact, the majority
of known global solutions in general relativity belong to this class.)
Pricisely this allowed one to give the physical interpretation to many
solutions known before only locally. That is, vacuum solutions to the
Einstein equations with cosmological constant from the considered class
describe black holes, wohmholes, cosmic strings, domain walls, and many
other solutions of physical interest.

The method of conformal blocks is applicable for two-dimensional
metrics having one Killing vector. In the present paper for an
arbitrary metric of this type the constructive rules for building
global solutions are formulated, smoothness and uniqueness of the
global solution are proved up to diffeomorphisms and the action of
descreet transformation group. In section \ref{sglocm} the metric is
considered, and conditions following from the singularity of the scalar
curvature are obtained. In section \ref{scoblo} the notion of the
conformal blocks is introduces from which global solutions are built.
In the next section \ref{sextsm} equations for extremals are solved,
their form, asymptotics, and completeness are analysed. In section
\ref{sglrul} the method of conformal blocks for building global solutions
is proposed, and the main theorem is formulated on the smoothness of
the obtained universal covering space. In section \ref{sexamp} the
Schwarzschild, Reisner--Nordstr\"om solutions, extremal and dilaton
black holes, and the constant curvature surfaces are considered as
examples of the conformal blocks method. In sections \ref{sedfic} and
\ref{sedpom} the proof of the main theorem is completed by the
transition to coordinate systems covering horizons and saddle points.
%*********************************************************************
\section{Local form of the metric                      \label{sglocm}}
%*********************************************************************
Consider a plane $\MR^2$ with Cartesian coordinates
$\z^\al$$=\left\{\tau,\s\right\}$, $\al=0,1$. Let conformally flat metric
of Lorentz signature be given in some domain in the plane
\begin{equation}                                        \label{emetok}
  ds^2=|N(q)|(d\tau^2-d\s^2).
\end{equation}
We consider conformal factor $N(q)\in {\cal C}^l$, $l\ge2$ to be
$l$ times continuously differentiable function of one argument
$q\in\MR$ except for a finite number of singularities. Later the
smoothness class of the global metric arising after continuation of
(\ref{emetok}) will be shown to coinside with the smoothness of the
conformal factor $N(q)$ between singularities. Let the argument $q$
to depend on one coordinate only, and this dependence to be given by
the ordinary differenrtial equation
\begin{equation}                                        \label{eshiff}
  \left|\frac{dq}{d\z}\right|=\pm N(q),
\end{equation}
with the following sign rule
\begin{equation}                                        \label{esignr}
\begin{array}{ccl}
N>0: & ~~\z=\s,   &~~\text{sign $+$, (static solution)},\\
N<0: & ~~\z=\tau, &~~\text{sign $-$, (homogeneous solution)}.
\end{array}
\end{equation}
The choice of the quadratic form (\ref{emetok}), (\ref{eshiff}) will
be shown in section \ref{sexamp} to include a wide class of metrics
interesting from mathematical and physical point of view. We admit that
conformal factor may have zeroes and singularities in a finite number
of points $q_i$, $i=1,\dots,k$. The infinite points $q_1=-\infty$ and
$q_k=\infty$ are also included in this sequence. We consider power
behaviour of the conformal factor near $q_i$
\begin{eqnarray}                                        \label{ecfapo}
  |q_i|<\infty:&~~~~~&N(q)\sim(q-q_i)^m,
\\                                                      \label{ecfasi}
  |q_i|=\infty:&~~~~~&N(q)\sim q^m.
\end{eqnarray}
The exponent $m$ may be an arbitrary real number, that is, the conformal
factor may have, for example, branch points. At finite points
$|q_i|<\infty$ for positive $m>0$ the conformal factor equals zero.
These points will be shown to correspond to {\em horizons} \index{Horizon}
of a space-time. Negative values $m<0$ correspond to singularities.
At infinite points positive and negative values of $m$ correspond
conversly to singularities and zeroes of the conformal factor.
The following analisys may be generalized to more complicated functions
$N(q)$ having, for example, logarithmic branch points. For simplicity we
restrict ouselves to a power behavior (\ref{ecfapo}). On the intervals
between zeroes and singularities where the function $N$ is either positive
or negative the solution is static with the Killing vector
$K=\pl_\tau$ or homogeneous with the Killing vector $K=\pl_\s$,
correspondingly. The length of the Killing vector equals $N(q)$ in
both cases.

Note that under the conformal transformations the form of the metric
(\ref{emetok}), (\ref{eshiff}) changes because the argument $q$ becomes
a function of both coordinates on the plane. Under the scale
transformations $\z'=k\z$, $k=\const\ne0$, which form a subgroup of
the conformal group the metric preserve its form if $q'=k^{-1}q$ and
$N'=k^{-2}N$.

In fact formulae (\ref{emetok}) and (\ref{eshiff}) define four
different metrics, due to the modulus sign on the derivative in
(\ref{eshiff}) there are two domains with static metric and two domains
with homogeneous metric differing by the sign of the derivative $dq/d\z$.
Denote these domains by the Roman digits
\begin{equation}                                        \label{emetdo}
\begin{array}{rcl}
  \text{I}:  ~~& N>0,~~& dq/d\s>0,\\
  \text{II}: ~~& N<0,~~& dq/d\tau<0,\\
  \text{III}:~~& N>0,~~& dq/d\s<0,\\
  \text{IV}: ~~& N<0,~~& dq/d\tau>0.
\end{array}
\end{equation}
The order of the domains will be clear from the following considerations.
Note that the static solution in the domain {\rm III} is obtained by the
space reflection $\s\rightarrow-\s$ from that in the domain {\rm I}, and
homogeneous solution in the domain {\rm IV} is related by time reflection
$\tau\rightarrow-\tau$ to the solution in the domain {\rm II}.
As far as changing the sign of the conformal factor may be compensated by
exchanging space and time coordinates $\tau\leftrightarrow\s$, then
homogeneous solutions in domains {\rm II} and {\rm IV} may be obtained
from static solutions for the conformal factor $-N$ in domains
{\rm III} and {\rm I} with the following rotation of the $\tau,\s$ plain
by the angle $\pi/2$ clockwise.

The Schwarzschild metric is an example of the type (\ref{emetok}).
Indeed, in static domains {\rm I} and {\rm III} make the coordinate
transformation $\tau,\s$ $\rightarrow$ $\tau,q$. Then the metric
takes the form
\begin{equation}                                        \label{eschco}
  ds^2=N(q)d\tau^2-\frac{dq^2}{N(q)},~~~~~~N>0.
\end{equation}
Neglecting the angular part in the Schwarzschild metric one gets the
metric (\ref{eschco}) for
\begin{equation}                                        \label{eschcf}
  N(q)=1-\frac{2M}q,
\end{equation}
where $M=\const$ is the mass of the black hole, and the coordinate $q$
may be interpreted as the radius. In the considered case the function
$N(q)$ has simple pole at $q=0$ and simple zero (horizon) at the point
$q=2M$. In the homogeneous domains {\rm II} and {\rm IV} one can make
anologous transformation of the coordinates $\tau,\s$ $\rightarrow$ $q,\s$.
Then the metric becomes
\begin{equation}                                        \label{eschho}
  ds^2=-\frac{dq^2}{N(q)}+N(q)d\s^2,~~~~~~N<0.
\end{equation}
In these domains the coordinate $q$ is timelike and can not be interpreted
as the radius. Coordinates $\tau,q$ and $q,\s$ are called
{\em Schwarzschild coordinates}. \index{Schwarzschild coordinates}
Metric in these coordinates has a simple form, all functions being given
explicitly. The disadvantage of the coordinates is that they do not
distinguish domains {\rm I, III} and {\rm II, IV}. This is essential,
because all domains must be used for the construction of a global solution.

Christoffel's symbols for the metric (\ref{emetok}) are different in
different domains. Explicit expressions for nonzero components only are
\begin{eqnarray}                                        \label{echrso}
\text{I}: &~~&
  \G_{\tau\tau}{}^\s=\G_{\tau\s}{}^\tau
  =\G_{\s\tau}{}^\tau=\G_{\s\s}{}^\s=\frac12N',\\
\text{II}: &~~&                                         \label{echrss}
  \G_{\tau\tau}{}^\tau=\G_{\tau\s}{}^\s
  =\G_{\s\tau}{}^\s=\G_{\s\s}{}^\tau=\frac12N',\\
\text{III}: &~~&                                        \label{echrst}
  \G_{\tau\tau}{}^\s=\G_{\tau\s}{}^\tau
  =\G_{\s\tau}{}^\tau=\G_{\s\s}{}^\s=-\frac12N',\\
\text{IV}: &~~&                                         \label{echrsf}
  \G_{\tau\tau}{}^\tau=\G_{\tau\s}{}^\s
  =\G_{\s\tau}{}^\s=\G_{\s\s}{}^\tau=-\frac12N',
\end{eqnarray}
where prime denotes the derivative with respect to $q$. In static
and homogeneous domains nonzero Christoffel's symbols have odd number
of space and time indices, respectively. In static and homogeneous
domains Christoffel's symbols differ by the sign. This is the consequence
of the fact that domains of one type are related by the transformation
$\z\rightarrow-\z$, and Christoffel's symbols are linear in derivatives.
The curvature tensor has only four nonzero components in every domain
\begin{eqnarray}                                        \label{ecurtc}
  \text{I,~III}: &~~&
  R_{\tau\s\tau}{}^\s=-R_{\s\tau\tau}{}^\s
  =R_{\tau\s\s}{}^\tau=-R_{\s\tau\s}{}^\tau=-\frac12N''N,
\\                                                      \label{ecurtt}
  \text{II,~IV}: &~~&
  R_{\tau\s\tau}{}^\s=-R_{\s\tau\tau}{}^\s
  =R_{\tau\s\s}{}^\tau=-R_{\s\tau\s}{}^\tau=~~\frac12N''N.
\end{eqnarray}
The Ricci tensor has two nonvanishing components
\begin{eqnarray}                                        \label{erictc}
  \text{I,~III}: &~~&
  R_{\tau\tau}=-R_{\s\s}=-\frac12N''N,
\\                                                      \label{erictt}
  \text{II,~IV}: &~~&
  R_{\tau\tau}=-R_{\s\s}=\frac12N''N.
\end{eqnarray}
And the scalar curvature is the same in all four domains
\begin{equation}                                        \label{esctmk}
  \text{I,~II,~III,~IV}:~~R=-N''.
\end{equation}
This is an algebraic equation relating $q$ and $R$ for a given
conformal factor $N$. As the consequence one gets that the function
$q=q(\tau,\s)$, as long as the scalar curvature may be considered as a
scalar function. That is equation (\ref{esctmk}) can be considered as an
invariant definition of the function $q(\z)$ in an arbitrary coordinate
system. In the domain where equation (\ref{esctmk}) can be solved with
respect to $q$, the scalar curvature $R$ may be chosen as one of the
coordinates.

Maximal extension means that if a surface has a boundary lying at a
finite distance, that is corresponding to a finite value of the
canonical parameter, then the boundary may be only singular where
the scalar curvature becomes infinite. In the opposite case a manifold
could be continued. For constant curvature surfaces to be discussed
in detail in section \ref{scocus} the conformal factor is a quadratic
polynomial
\begin{equation}                                        \label{ecocsp}
  N(q)=-(aq^2+bq+c),~~~~~~a,b,c=\const.
\end{equation}
Here $R=2a$.

As the consequence of equation (\ref{esctmk}) the scalar curvature
is singular near $q_i$ for the following exponents in asymptotic
behaviour (\ref{ecfapo}):
\begin{eqnarray}                                        \label{esfpsc}
  |q_i|<\infty:&~~~~~&m<0,~~0<m<1,~~1<m<2,
\\                                                      \label{esfpcc}
  |q_i|=\infty:&~~~~~&m>2.
\end{eqnarray}
At infinite points $q\rightarrow\pm\infty$ the scalar curvature tends
to a nonzero constant for $m=2$ and to zero for $m<2$.
%*********************************************************************
\section{Conformal blocks                              \label{scoblo}}
%*********************************************************************
The notion of the conformal block corresponding to every interval
$(q_i,q_{i+1})$ will be used in building of global solutions.
Inside the interval the conformal factor $N$ is either strictly
positive or strictly negative. Let us find the domain of definition
for the metric (\ref{emetok}) on the $\tau,\s$ plain. For definiteness
we consider static solution {\rm I} on the interval $(q_i,q_{i+1})$.
Then the time coordinate takes values on the whole real axis
$\tau\in\MR$. Equation (\ref{eshiff}) for the space coordinate $\s$
becomes
\begin{equation}                                        \label{estsos}
  \frac{dq}{d\s}=N(q).
\end{equation}
Integration constant for this equation corresponds to a shift of the
spacelike coordinate $\s\rightarrow\s+\const$, that is to the choice
of the origin of $\s$. From equation (\ref{estsos}) one gets that the
domain of $\s$ depends on the convergence of the integral
\begin{equation}                                        \label{eforin}
  \s\sim\int^{q_i,q_{i+1}}\frac{dq}{N(q)}
\end{equation}
at the boundary points. The integral (\ref{eforin}) converge or diverge
depending on the exponent $m$:
\begin{equation}                                        \label{eboucb}
\begin{array}{rl}
  |q_i|<\infty:~~~~&\left\{
  \begin{array}{rll}
  m<1  ~~&\text{-- converge,}  &\text{~~line,}\\
  m\ge1~~&\text{-- diverge,}&\text{~~angle,}
  \end{array}\right. \\
  |q_i|=\infty:~~~~&\left\{
  \begin{array}{rll}
  m\le1~~&\text{-- diverge,}&\text{~~angle,}\\
  m>1  ~~&\text{-- converge,}  &\text{~~line.}
  \end{array}\right.
\end{array}
\end{equation}
At the right of this table the form of the boundary of the corresponding
conformal blocks introduced below is given. If at both ends of the
interval $(q_i,q_{i+1})$ the integral diverge, then
$\s\in(-\infty,\infty)$, and the metric is defined on the whole
$\tau,\s$ plain. If at one of the ends $q_{i+1}$ or $q_i$ the integral
converge, then the metric is defined on the half plain
$\s\in(-\infty,\s_{i+1})$ or $\s\in(\s_i,\infty)$, correspondingly.
The choice of boundary points $\s_{i+1}$ and $\s_i$ is arbitrary,
and without loss of generality one may set $\s_{i,i+1}=0$.
If the integral converge on both sides of the interval, then the
solution is defined on the strip $\s\in(\s_i,\s_{i+1})$, and one point
can be only set to zero.

\begin{figure}[htb]%-------------------------------------------------
 \begin{center}
 \leavevmode
 \epsfxsize=120mm
 \epsfbox{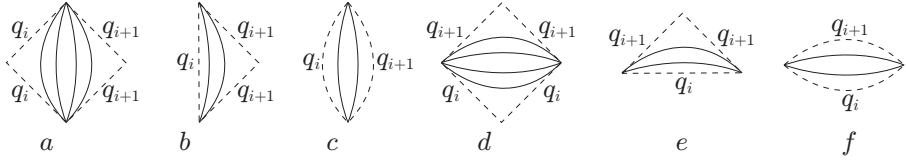}
 \end{center}
 \caption{ Conformal blocks for static ({\it a,b,c}) and homogeneous
 ({\it d,e,f}) solutions. Thin solid lines denote Killing trajectories.
 The variable $q\in(q_i,q_{i+1})$ is constant along Killing trajectories
 and varies from the left boundary to the right one being constant along the
 boundaries. For homogeneous solutions the variable $q$ varies
 analogously but from bottom to top. Every conformal block has its
 symmetrical patner obtained by space $\s\leftrightarrow-\s$ or time
 $\tau\leftrightarrow-\tau$ reflection for static and homogeneous
 solutions, respectively.\label{fcoblo}}
\end{figure}

For visual image of a maximally extended solution the notion of
conformal block is introduced. To this end we map the $\tau,\s$
plain on the square along the light like directions
\begin{equation}                                        \label{elicco}
  \x=\tau+\s,~~~~~~\eta=\tau-\s.
\end{equation}
For the purpose we make a conformal transformation
\begin{equation}                                        \label{etrfun}
  u=u(\x),~~~~v=v(\eta),~~~~~~u,v\in{\cal C}^{l+1},
\end{equation}
where the functions $u$ and $v$ are bounded, and their first derivatives
are positive. Without loss of generality we consider $u,v\in[-1,1]$.
The assumption about the smoothness of the transition functions
preserves the smoothness of the metric after the coordinate
transformation. Then the static solution defined on the whole
$\tau,\s$ plain corresponds to a square {\em conformal block},
\index{Conformal block}\index{Block conformal}%
shown in Fig.~\ref{fcoblo},{\it a}.
%---------------------------------------------------------

In a general case conformal block is a finite part of the $u,v$ plain
and the given metric (\ref{emetok}) in coordinates (\ref{elicco}),
(\ref{etrfun}). If one requires the extremals to approach the boundary
on all possible angles, then this uniquely determines the asymptotic
behavior of the transition functions (\ref{etrfun}) \cite{Katana93A}.
In all other respects the transition functions are arbitrary.
We shall not discuss this point in detail because the smoothness of the
metric on the horizons will be proved by going to Eddington--Finkelstein
coordinates in section \ref{sedfic}.

The value of $q$ and, consequently, the scalar curvature is constant
along time like Killing trajectories shown inside the block by thin
solid lines.
The variable $q$ continuoesly changes from left to right monotonically
increasing in the domain {\rm I} and decreasing in the domain {\rm III}.
It takes the value $q_i$ on both left sides of the square and the value
$q_{i+1}$ on both right sides. Lower and upper coners of the square are
essentially singular points because the limit of $q$ in these points
depends on the path along which a sequence goes to the coner. We say
that static conformal block has two boundaries, left and right, on which
the variable $q$ takes values $q_i$ and $q_{i+1}$, respectively.

If solution of equation (\ref{estsos}) is defined on the half interval,
then a static solution corresponds to a triangular conformal block,
an example of which is shown in Fig.~\ref{fcoblo},{\it b}. And the
vertical line correspond to the boundary point $\s_i=0$. This can be
always achieved by the choice of the functions $u(\x)$ and $v(\eta)$.

When solutions of equation (\ref{estsos}) is defined on a finite
interval the conformal block is represented in the form of a lens,
depicted in Fig.~\ref{fcoblo},{\it c}. The choice of the functions
$u(\x)$ and $v(\eta)$ allows one to make left or right boundary
vertical but not both boundaries simultaneously.

Let us remind the reader that there are two conformal blocks for
every interval because equation (\ref{eshiff}) is invariant under the
space reflection $\s\leftrightarrow-\s$ due to the modulus sign.

Analogously, one of the conformal blocks shown in
Fig.~\ref{fcoblo},{\it d,e,f} or its turned over partner obtained
by the time inversion $\tau\leftrightarrow-\tau$ corresponds to every
homogeneous solution. On these conformal blocks the variable $q$ is
constant along space like Killing trajectories and monotonically
increases or decreases from the value $q_i$ on the lower boundary to
the value $q_{i+1}$ on the upper one. Left and right coners of
homogeneous conformal blocks are essentially singular points.

Conformal blocks for static and homogeneous solutions are called
static and homogeneous, respectively.

Thus we have shown that a Lorenzian surface with a given conformally
flat metric (\ref{emetok}) is diffeomorphic to one of the conformal
blocks. To construct maximally extended solutions one has to find
and analyse extremals on these surfaces, in the case of necessity
to extend the manifold and the metric, and to prove the smoothness.
The use of the conformal blocks in construction of global solutions
is convenient because all light like extremals are depicted as two
classes of straight lines going by the angle $\pm\pi/4$ to the time
axis. Therefore gluing conformal blocks along edges preserves the
smoothness of light like extremals.
%*********************************************************************
\section{Extremals                                     \label{sextsm}}
%*********************************************************************
%*********************************************************************
\subsection{The form of the extremals                  \label{sextfo}}
%*********************************************************************
To understand the global structure of a solution for metric (\ref{emetok})
one has to analyse the behavior of extremals
$\z^\al(t)=\left\{\tau(t),\s(t)\right\}$ satisfying in general the
system of equations
$$
  \ddot\z^\al=-\G_{\bt\g}{}^\al\dot\z^\bt\dot\z^\g,
$$
where the dot denotes the derivative with respect to the canonical
parameter $t$.

Let us analyse the behavior of extremals in the static domain
\begin{equation}                                        \label{emetfd}
  \text{I}:~~ds^2=N(q)(d\tau^2-d\s^2),~~~~~~\frac{dq}{d\s}=N(q)>0.
\end{equation}
Using the expression for Christoffell's symbols (\ref{echrso}) we get
equations for extremals
\begin{eqnarray}                                        \label{eqexts}
  \ddot\tau &=& -N'\dot\tau\dot\s,
\\                                                      \label{eqextt}
  \ddot\s   &=& -\frac12N'(\dot\tau^2+\dot\s^2).
\end{eqnarray}
This system of equations has two first integrals
\begin{eqnarray}                                        \label{efisco}
  N(\dot\tau^2-\dot\s^2)&=&C_0=\const,
\\                                                      \label{efisct}
  N\dot\tau&=&C_1=\const.
\end{eqnarray}
The integral of motion (\ref{efisco}) has the same meaning as the
possibility to choose the length of an extremal as the canonical
parameter for a positive definite metric. The value of the constant
$C_0$ defines the type of an extremal:
\begin{eqnarray*}
  C_0>0 &-& \text{time like extremals},
\\
  C_0=0 &-& \text{light like extremals},
\\
  C_0<0 &-& \text{space like extremals}.
\end{eqnarray*}
Note that the type of an extremal does not change when going from
one point to another one. The second integral (\ref{efisct}) is the
consequence of the existence of a Killing vector field, that is with
the symmetry of the problem, and for an arbitrary metric it does not
exist.
\begin{Theorem}                                         \label{textrf}
Any extremal in the static space-time of the type {\rm I, III} belongs
to one of the following four classes.\newline
1) Light like extremals
\begin{equation}                                        \label{extlif}
  \tau=\pm\s+\const,
\end{equation}
with the canonical parameter defined by the equation
\begin{equation}                                        \label{extlip}
  \dot\tau=\frac1N.
\end{equation}
2) General type extremals the form of which is defined by the equation
\begin{equation}                                        \label{extgef}
  \frac{d\tau}{d\s}=\pm\frac1{\sqrt{1-C_2 N}},
\end{equation}
where $C_2$ is an arbitrary nonzero constant, the values of $C_2<0$ and
$C_2>0$ defining space and time like extremals, respectively. The canonical
parameter is defined by one of the two equations
\begin{eqnarray}                                        \label{extgpf}
  \dot\tau&=&\frac1N,
\\                                                      \label{extgps}
  \dot\s  &=&\pm\frac{\sqrt{1-C_2N}}N.
\end{eqnarray}
The plus or minus signs in equations (\ref{extgef}) and (\ref{extgps})
are to be chosen simultaneously.\newline
3) Straight space like extremals parallel to $\s$ axis and going through
every point $\tau=\const$. The canonical parameter is defined by the
equation
\begin{equation}                                        \label{extstp}
  \dot\s=\frac1{\sqrt N}.
\end{equation}
4) Straight degenerate time like extremals parallel to $\tau$ axis and
going through the critical points $\s_0=\const$ where
\begin{equation}                                        \label{extdcp}
  N'(\s_0)=0.
\end{equation}
Their canonical parameter coinsides with the time like coordinate
\begin{equation}                                        \label{extdgp}
  t=\tau.
\end{equation}
\end{Theorem}
{\bf Proof} of the theorem reduces to the integration of the system
of equations (\ref{eqexts}), (\ref{eqextt}). In obtaining the integral
of motion (\ref{efisct}) one has to devide on $\dot\tau$ and $\dot\s$,
therefore these degenerate cases must be considered seperately. For
$\tau=\const$ equation (\ref{eqexts}) is trivially fullfield, and
equation (\ref{eqextt}) after integration yeilds
$$
  -N\dot\s^2=C_0\ne0.
$$
This equations is reduced to (\ref{extstp}) by the scaling of the
canonical parameter.

The second degenerate case corresponds to $\dot\s=0$. Hence equation
(\ref{eqextt}) is fullfilled only in those points $\s=\s_0$ where
$N'(\s_0)=0$ and equation (\ref{eqexts}) reduces to the equation
$\ddot\tau=0$. Therefore the $\tau$ coordinate can be chosen as the
canonical parameter. One may note the validity of the equality
$$
  \frac{dN}{d\s}=N\frac{dN}{dq}=0,
$$
and differentiation with respect to $q$ in the equation (\ref{extdcp})
can be changed to the differentiation with respect $\s$ because $N\ne0$.
In this way all special cases are exhausted.

The remaining cases are obtained from the analysis of the integrals
(\ref{efisco}), (\ref{efisct}), the constant $C_0$ being arbitrary and
$C_1\ne0$, since the case $C_1=0$ is related to the extremals of the
third class. Equation (\ref{efisco}) with (\ref{efisct}) yeilds
\begin{equation}                                        \label{extsgt}
  \dot\s=\pm\frac{C_1}N\sqrt{1-C_2N},
\end{equation}
where
\begin{equation}                                        \label{econct}
  C_2=\frac{C_0}{C_1^2}.
\end{equation}
The equation for the form of a general type extremals (\ref{extgef}) is
the consequence of (\ref{efisct}) and (\ref{extsgt}), the value of the
constant $C_2$ defining the type of an extremal. The equation for the
canonical parameter is obtained from (\ref{efisct}) and (\ref{extsgt})
after rescaling. In a particular case $C_2=0$ one gets the light like
extremals (\ref{extlif}).
$\Box$

As the consequence of equation (\ref{extgef}) one gets that the constant
$C_2$ parametrizes the angle at which the extremal goes through a given
point. One may check that through every point in any direction goes one
and only one extremal.

\begin{figure}[htb]%-------------------------------------------------
 \begin{center}
 \leavevmode
 \epsfxsize=120mm
 \epsfbox{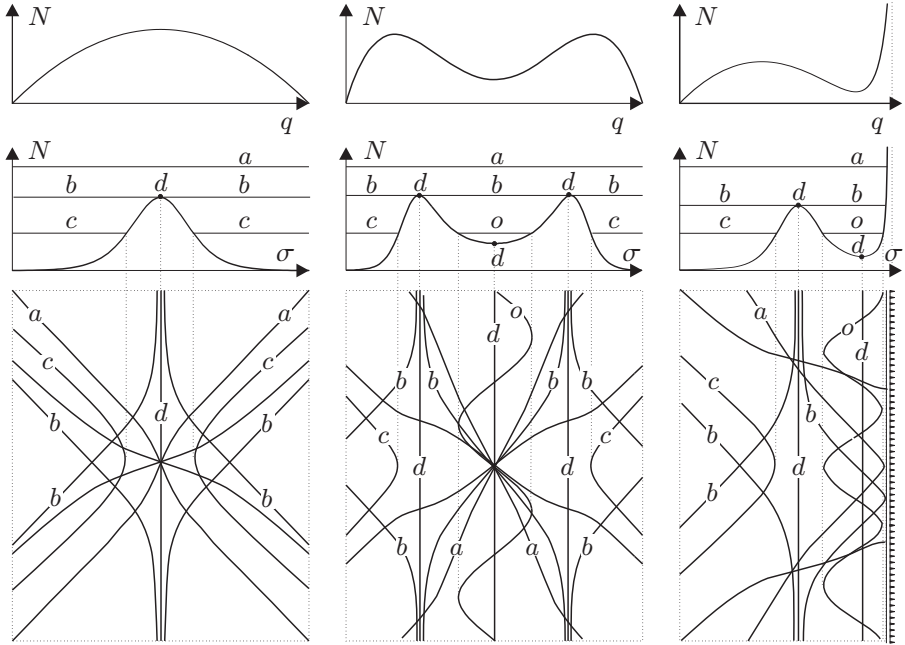}
 \end{center}
 \caption{The first raw: tipical behaviour of the conformal factor
 $N(q)$ between two zeroes and a zero and a singularity for a static
 solution. The second raw: the dependence of the conformal factor
 $N(\s)$ on the space like coordinate. The third raw: tipical time
 and space like extremals. The letters mark time like extremals
 having qualitatively different behaviour for different values of the
 constant $C_2$. Through any local extrema goes one degenerate extremal
 ({\it d}). There are oscillating extremals ({\it o}) in the second and
 third cases.
 \label{fextrema}}
\end{figure}

Tipical behaviour of the conformal factor with one and three local
extrema between two zeroes and with two local extrema between a zero
and a singularity for static conformal blocks is shown in
Fig.~\ref{fextrema} in the first row.
%---------------------------------------------------------
The dependence of the conformal factors on the space like coordinate is
shown in the second raw. The qualitative behaviour of time and space
like extremals. Not to overload the figure light like extremals going
through every point of the space-time are omitted. All extremals can
be shifted along the time like coordinate $\tau$.

Time like general type extremals have qualitatively different behaviour
for different values of the constant $C_2$. They are marked by the letters
{\it a,b,c} and {\it o}. If both boundary points $q_i$ and $q_{i+1}$ of the
interval are zeroes, then, as the consequence of continuity, the conformal
factor $N$ has at least one maximum through which goes the degenerate
extremal ({\it d}). For sufficiently large values of $C_2$ the extremals
cross local maxima ({\it a}). If the function $N$ has a local minimum
through which goes the degenerate extremal as well, then there are
oscillating extremals among general type extremals which oscillate
around the local minimum ({\it o}). This is the consequence of equation
(\ref{extgef}) because for positive values of $C_2$ the range of the
coordinate $\s$ is given by the inequality
$$
  N<\frac1{C_2}.
$$
Critical points $\s_*$ in which $N(\s_*)=1/C_2$ are the turning points
for a finite value $\tau=\tau_*$ if and only if the integral
\begin{equation}                                        \label{eintpo}
  \int^{\tau_*} d\tau=\pm\int^{\s_*}\frac{d\s}{\sqrt{1-C_2N}}
\end{equation}
converges. This integral converges if the point $\s_*$ does not coinside
with the local maximum. That is the turn happens for a finite value of
$\tau_*$ and the canonical parameter as the consequence of equation
(\ref{extgps}). Corresponding nonoscillating extremals are marked by the
letter ({\it c}). If the critical point coinsides with the local maximum,
then the integral (\ref{eintpo}) diverges. It means that the critical
point is reached at an infinite value of $\tau_*$ and the canonical
parameter. That is the corresponding extremal ({\it b}) is complete in
this direction.

The above theorem describes all extremals for static solutions. The
behaviour of extremals for homogeneous solutions is similar. They are
obtained by the replacement $N\rightarrow-N$, the exchange of time and
space like coordinates $\tau\leftrightarrow\s$ and by the rotation of
the whole plain by the angle of $\pi/2$.
%*********************************************************************
\subsection{Asymptotics of the extremals               \label{sextac}}
%*********************************************************************
To solve the problem whether a solution defined on the $\tau,\s$ plain
is complete or it must be continued one has to analyze asymptotics and
completeness of extremals near the boundary of a conformal block.
All time like extremals for which
$$
  \lim_{\tau\to\pm\infty}\left|\frac{d\tau}{d\s}\right|>1
$$
go to upper and lower corners of a conformal block if they exist.
Extremals for which
$$
  \lim_{\s\to\pm\infty}\left|\frac{d\tau}{d\s}\right|<1
$$
go to left and right corners of a conformal blaocks. The extremals with
a light like asymptotics
$$
  \lim_{\tau,\s\to\pm\infty}\left|\frac{d\tau}{d\s}\right|=1
$$
go to an edge of a square conformal block.

Let us consider asymptotics of extremals near the boundary of static
conformal blocks. Any light like extremal starts and ends at the
opposite edges of a square conformal block. On triangular or lens
type conformal blocks a light like extremal starts or ends on a time
like boundaries.

For static solutions {\rm I, III} straight extremals parallel to the
space like $\s$ axis start in the left and end in the right corner
of a square conformal block. On a triangular or lens type conformal
block these extremals start or (and) end on a time like boundary.
Degenerate extremals and general type oscillating ones start in the
lower and end in the upper corner of a conformal block.

Let us consider nonoscillating general type extremals. If a boundary
of a conformal block is a horizon, $N(q_i)=0$, $|q_i|<\infty$, then
tangent vectors to extremals (\ref{extgef}) have light like
asimptotic behaviour
$$
  \lim_{q\to q_i}\left|\frac{d\tau}{d\s}\right|=1.
$$
And these general type extremals go to edges of a square and not
to the corners. Indeed, for extremals going to the right upper edge
of a square equations (\ref{extgpf}), (\ref{extgps}) imply
$$
  \frac{d\eta}{d\x}=\frac{1-\sqrt{1-C_2N}}{1+\sqrt{1+C_2N}}.
$$
As the consequense the integral
$$
  \int d\eta\sim\int d\x N\sim\int^{q_i}dq=q_i
$$
converges, and this corresponds to a point on the edge of a square.
If time like boundary is reached at finite values of $\s_i$, then
a tangent vector to a space like extremal has the asimptotics
$$
  \frac{d\tau}{d\s}\sim\frac1{\sqrt N}\to0.
$$
That is they reach the boundary $|\s_i|<\infty$ at the right angle.
Time like extremals can not reach a time like boundary because the
right hand side in (\ref{extgef}) for $C_2>0$ becomes imaginary
near the singularity. These extremals near the boundary $|\s_i|<\infty$
have a turning point for finite values of $\tau$ for the same reason
as the oscillating extremals.
%*********************************************************************
\subsection{Completeness of extremals                  \label{sextсо}}
%*********************************************************************
Theorem \ref{textrf} allows one to analyze the completeness of extremals
near the boundary of a conformal block. For definiteness we consider a
static solution. First of all note that degenerate extremals are always
complete because their canonical parameter coinsides with the time
(\ref{extdgp}). Oscillating extremals are complete as well because they
make infinite number of oscillations each of them corresponding to a
finite variation of a canonical parameter. If a general type extremal
for $\tau\to\pm\infty$ approach the degenerate extremal then it is
complete because equation (\ref{extgpf}) for the canonical parameter
imply
$$
  \lim_{\s\to\s_0}t\to\int^{\infty}d\tau N(\s_0)\to\infty.
$$
For a stationary solution in lower and upper corners of a conformal
block go only degenerate and oscillating extremals, hence they are
always complete. The absence of these extremals is possible. In this
case we consider these corners as complete because any time like
curve has infinite length at $\tau\to\pm\infty$. This means that
lower and upper corners of a stationary conformal blocks are always
complete, that is they are past and future time infinities,
correspondingly.

Completeness of light like extremals is defined by the equation
(\ref{extlip}) which imply
\begin{equation}                                        \label{elieco}
  \lim_{\s\to\pm\infty}t\to\int d\s N=\int^{q_i}dq=q_i.
\end{equation}
This means that near the boundary of a conformal block they are
incomplete for finite $q_i$ and complete for $|q_i|=\infty$.

Completeness of nonoscillating general type extremals going to the
boundary corresponding to $q_i$ follows from the equation
(\ref{extgps})
\begin{equation}                                        \label{extgtc}
  \lim_{q\to q_i}t\to\int d\s\frac N{\sqrt{1-C_2N}}
  \sim\int^{q_i}\frac{dq}{\sqrt{1-C_2N}}.
\end{equation}
Near horizons $N\rightarrow0$ behaviour of general type extremals
is the same as of light like extremals (\ref{elieco}). The singularity
$N\rightarrow\infty$ is reached by space like extremals, $C_2<0$,
their completeness being defined by the equation
\begin{equation}                                        \label{extgpc}
  \lim_{q\to q_i}t\to\int^{q_i}\frac{dq}{\sqrt{N}}.
\end{equation}
These extremals are incomplete at finite points $|q_i|<\infty$ for $m<0$.
At infinite points $|q_i|=\infty$ space like extremals are complete for
$0<m\le2$ and incomplete for $m>2$.

Completeness of straight extremals parallel to $\s$ axis is defined
by equation (\ref{extstp}) and near the boundary $q_i$ is given by the
integral (\ref{extgpc}). This means that their completeness near
singularities is the same as for general type extremals. Near zeroes
they are always complete except for the case of a simple zero at a
finite point where they are incomplete.

\begin{figure}[htb]%-------------------------------------------------
 \begin{center}
 \leavevmode
 \epsfxsize=120mm
 \epsfbox{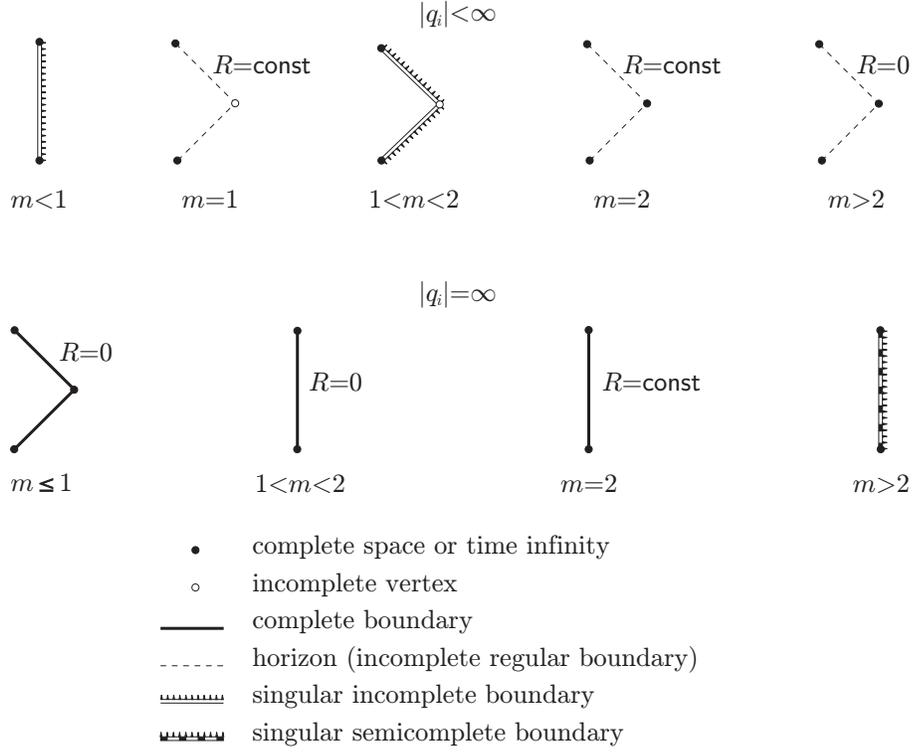}
 \end{center}
 \caption{The form of the right boundary of the static conformal block
 depending on the exponent $m$. In upper and lower raws the boundaries
 are shown for finite and infinite values of $q_i$, respectively.
 \label{fboupr}}
\end{figure}

Thus completeness of all boundary elements is analysed. The summary
is given in Fig.~\ref{fboupr}.
%---------------------------------------------------------
Here boundary elements of conformal blocks are shown depending on
the exponent $m$ (\ref{ecfapo}) for finite and infinite values of $q_i$.
For definiteness we show the right boundary of static conformal blocks
{\rm I}. In the figure time like boundary is shown by a vertical line
and light like boundary by an angle. If the scalar curvature is singular
on the boundary (\ref{esfpsc}), (\ref{esfpcc}), then it has protuberances.
Incomplete and complete boundaries are shown by dashed and thick solid lines,
respectively. Exclusion is the semicomplete boundary corresponding to
$|q_i|=\infty$ for $m>2$, Fig.~\ref{fboupr}. Light like extremals reaching
this boundary are complete while space like ones are not. Lower and upper
corners of all static conformal blocks (time past and future infinities)
are essentially singular poits and always complete, the completeness is
shown by filled circles. Completeness and incompleteness of right corners
of angle boundaries are shown by filled and not filled circles,
respectively. Left boundaries have the same structure, but angles should
be reflected with respect to a vertical line.

Qualitative behaviour of all extremals in homogeneous space-time
{\rm II, IV} is defined by zeroes and singularities of the conformal
factor in the same way as for a static solution. Corresponding conformal
blocks should be turned by the angle $\pi/2$.
%*********************************************************************
\section{Construction of global solutions              \label{sglrul}}
%*********************************************************************
In section \ref{scoblo} the conformal block was associated to every
solution of type {\rm I-IV} defined on the interval $(q_i,q_{i+1})$.
Then in section \ref{sextсо} the completeness or incompleteness of
their boundaries was proved by the analysis of extremals. As a result
horizons and only them turned out to correspond to boundaries incomplete
with respect to extremals, the scalar curvature being finite on them.
In all other cases the boundary is either complete or the curvature is
singular. Therefore solutions of the form (\ref{emetok}) must be
continued only through horizons, that is the zeroes of the conformal
factor. Global solutions will be depicted with the help of
Carter--Penrose diagrams. Carter--Penrose diagram is a bounded image
of maximally extended surface in which two classes of light like
extremals are depicted as two classes of perpendicular straight lines
on the Minkowskian plain. Let us formulate the rules by which maximal
extension of a manifold with metric (\ref{emetok}) for a given
conformal factor $N(q)$ is realised. Smoothness and uniqueness of the
manifold constructed following these rules is given by the theorem
\ref{tunisp}.
\begin{enumerate}
\item Every global solution for a metric (\ref{emetok}) corresond to the
interval of the variable $q\in(q_-,q_+)$, where $q_\pm$ are either
infinite points or a curvature singularities defined by the condition
(\ref{esfpsc}). Singularities inside the interval must be absent.
\item If there are no zeroes inside the interval $(q_-,q_+)$ then the
corresponding conformal block is the maximally extended solution.
\item If there are zeroes (horizons) inside the interval $(q_-,q_+)$
then enumerate them, $N(q_j)=0$, $j=1,\dots,n$, $q_j\in(q_-,q_+)$,
and associate with each of the interval $(q_-,q_1),\dots,(q_n,q_+)$
a pare of static or homogeneous conformal blocks for $N>0$ и $N<0$,
respectively.
\item Sew together conformal blocks along horizons $q_j$, preserving the
smoothness of the conformal factor, that is sew together conformal blocks
corresponding only to adjacent intervals $(q_{j-1},q_j)$ and
$(q_j,q_{j+1})$, and if the gluing is performed for static or
homogeneous conformal blocks, then sew together blocks of one type.
\item The Carter--Penrose diagram obtained by gluing all adjacent
conformal blocks is a connected fundamental region if inside the
interval $(q_-,q_+)$ the conformal factor changes its sign.
If $N\ge0$ or $N\le0$ everywhere inside the interval $(q_-,q_+)$
then one gets two fundamental regions related by space or time
reflection.
\item For one zero of an odd degree the boundary of the fundamental
region consists of boundaries of the conformal blocks corresponding
to the points $q_-$ and $q_+$, and the Carter--Penrose diagram
represents the global solution.
\item If there is one zero of even degree or two or more zeroes of
arbitrary degree the boundary of the fundamental region includes
horizons, and it has to be either continued periodically in space
and (or) time or the opposite sides should be identified.
\item If the fundamental group of the Carter--Penrose diagram is
trivial then it is the universal covering space for a global
solution.
\item If the fundamental group of the Carter--Penrose diagram is
nontrivial, then construct the corresponding universal covering
space.
\end{enumerate}
The statement 1) is the consequence of impossibility to continue
the solution through the points $q_\pm$ because these points are
either complete or singular. Solutions are extended only through
horizons $|q_j|<\infty$ with $m=1$ or $m\ge2$. In these cases the
boundary of a conformal block is an angle. If the value $q_j$
corresponds to an odd zero, then all four conformal blocks for the
intervals $(q_{j-1},q_j)$ and $(q_j,q_{j+1})$ are sewed together according
to the rules 3) and 4) around the vertex $q_j$, this point being the
saddle point. If $q_j$ is a zero of even degree then the regions of
one type are only sewed together along horizons. If inside the interval
$(q_-,q_+)$ the function $N$ does not change the sign, then two
disconnected fundamental regions appear, each of them can be periodically
continued. When $N$ changes its sign
the saddle point appear, and regions of different types form one
fundamental region which can be either periodically continued or not
depending on the value of $q$ on the boundary. This is the meaning of
the rules 5) and 6). Let us formulate the main theorem justifying the
above mentioned rules for construction of global solutions.
\begin{Theorem}                                         \label{tunisp}
The universal covering space constructed according to the rules 1)--9)
is the maximally extended pseudo riemannian $\CC^{l+1}$ manifold with
the continuous ${\cal C}^l$, $l\ge2$, metric such that every point not
lying on the horizon has a neighbourhood diffeomorphic to some domain
with the metric (\ref{emetok}).
\end{Theorem}
{\bf Proof.} By construction the interiour of a conformal block is
covered by one chart and is a ${\cal C}^{l+1}$ manifold as the
consequence of (\ref{etrfun}). Therefore inside a conformal block
the metric is of the same class as the conformal factor. The
transition functions (\ref{etrfun}) isometrically map the $\tau,\s$
plain on the interior of the conformal block. Therefore one has to
prove the smoothness of manifold and metric on the horizons and
saddle points only. This is done in sections \ref{sedfic} and
\ref{sedpom}, respectively. The uniqueness of the resulting manifold
follows from the well known theorem from algebraic topology
stating that every manifold with nontrivial fundamental group has
a unuque universal covering space up to diffeomorphisms
(see, for example, \cite{DuNoFo86}).
$\square$

Before concluding the proof of the theorem we consider several examples
to clear how to use the rules for construction of global solutions, and
what kind of solutions may appear.
%Заметим, что универсальное накрывающее пространство всегда является
%ориентируемым. Ниже будет показано, каким образом можно получить
%неориентируемые гладкие многообразия при склеивании конформных блоков.
%*********************************************************************
\section{Examples                                      \label{sexamp}}
%*********************************************************************
If the metric of the form (\ref{emetok}) is obtained as the result of
solution of some equations of motion, then following the rules 1)--9)
from the preceeding section one can construct the global solution
do not worring about going to new coordinate systems covering larger
domains. The advantage of the method is that elementary analysis of
the conformal factor $N(q)$ is sufficient for construction of global
solutions. We start from three well known examples from general
relativity.
%*********************************************************************
\subsection{Schwarzschild solution                     \label{schwas}}
\begin{figure}[htb]%-------------------------------------------------
 \begin{center}
 \leavevmode
 \epsfxsize=120mm
 \epsfbox{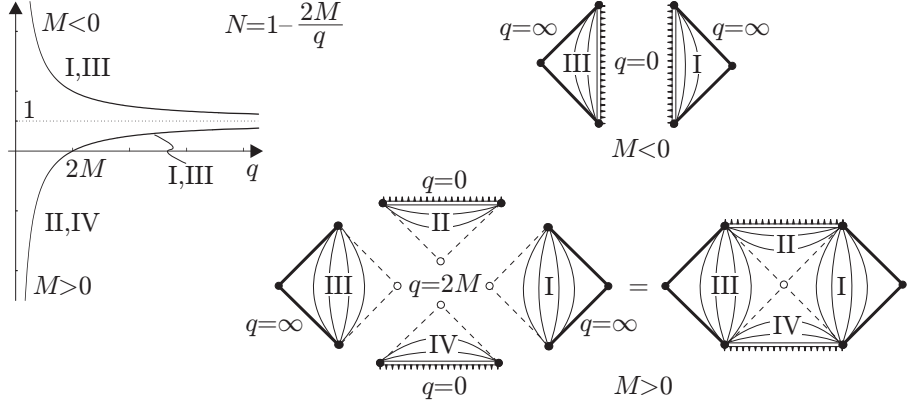}
 \end{center}
 \caption{The behavior of the conformal factor for the Schwarzschild
 solution for positive and negative mass. For $M<0$ there are two
 global solutions depicted by triangular Carter--Penrose diagrams.
 For $M>0$ four conformal blocks corresponding to two intervals
 $(0,2M)$ и $(2M,\infty)$ are sewed together in one global solution.
 \label{fschws}}
\end{figure}

%*********************************************************************
The conformal factor for the Schwarzschild solution has the form
(\ref{eschcf}) with $M>0$. It has a simple pole at $q=0$, $m=-1$, which
corresponds to curvature singularity (\ref{esfpsc}). The point $q_1=2M$,
$m=1$, is a simple zero and corresponds to a horizon. The values
$q=\pm\infty$, $m=0$, correspond to asymptotically flat space infinity.
The behaviour of the conformal factor is shown in Fig.~\ref{fschws}.
%---------------------------------------------------------
We see that two global solutions for positive and negative $q$
correspond to the infinite interval $q\in(-\infty,\infty)$. For
positive $q\in(0,\infty)$ one has $q_-=0$, $q_1=2M$ and $q_+=\infty$.
For each interval $q\in(0,2M)$ and $q\in(2M,\infty)$ there are two
homogeneous and static conformal blocks depicted in Fig.~\ref{fschws}.
The boundary elements are defined in Fig.~\ref{fboupr}. The global
solution shown in Fig.~\ref{fschws} is uniquely constructed by gluing
together these four conformal blocks according to the rule 4). This
Carter--Penrose diagram represents the Kruskal--Szekeres extention of the
Schwartzschild solution. Let us remind that the Kruskal--Szekeres
extention is the Schwarzschild metric written in such a coordinate
system which covers all domains {\rm I--IV}. Carter was the first who
depicted this extension as the bounded region on the plain \cite{Carter73}.

The constructed global solution is the unique, up to diffeomorphisms,
universal covering space because its fundamental group is trivial,
and part of it, namely, the domain {\rm I} or {\rm III} is diffeomorphic
to the Schwarzschild solution.

Similar global solutions one will have for a wide class of metrics
with the conformal factor having qualitatively the same form as the
lower branch in Fig.~\ref{fschws}. That is the conformal factor is
defined on an infinite half-interval $(q_-,\infty)$, has singularity
at $q_-$, one zero, and goes to a constant at infinity. The theorem
\ref{tunisp} quarantees the uniqueness and smoothness of the global
solution, and one does not need to find global coordinates explicitly.

Note that the surface represented by the Carter--Penrose diagram has
nonconstant scalar curvature
\begin{equation}                                        \label{escshs}
  R=\frac{4M}{q^3}.
\end{equation}
This is the two-dimensional scalar curvature on the Lorentz surface.
The scalar curvature for a four-dimensional metric with the angular
dependence is identically zero due to Einstein's equations. Note that
the two-dimensional scalar curvature (\ref{escshs}) coinsides with
the invariant eigen value of four dimensional Weyl tensor \cite{LanLif62}.
The center of the Carter--Penrose diagram is the saddle point for
variable $q$ and, consequently, for the scalar curvature (\ref{escshs}).
This point is incomplete and denoted by the unfilled circle.

For negative values of $q$ horizons are absent.
Therefore triangular conformal blocks shown in Fig.~\ref{fschws}
represent maximally extended solutions. In this space-time the curvature
singularity is along the time like boundary lying at a finite distance
and is not surrounded by a horizon. Singularities of this type are
called naked. Global solution for negative $q$ can be considered as
a global solution for positive $q$ allowing one to interpret the
coordinate $q$ as the radius but with negative mass $M<0$. These
solutions are considered as unphysical.
%*********************************************************************
\subsection{Reissner--Nordstr\"om solution             \label{srenos}}
\begin{figure}[htb]%-------------------------------------------------
 \begin{center}
 \leavevmode
 \epsfxsize=120mm
 \epsfbox{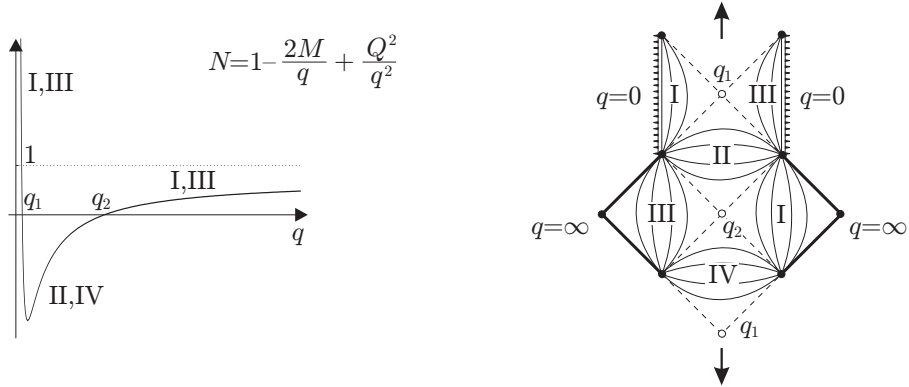}
 \end{center}
 \caption{The conformal factor and the fundamental region for the
 Reissner--Nordstr\"om solution. Arrows show possible extension or
 identification of the solution along horizonz.\label{frenpd}}
\end{figure}

%*********************************************************************
The conformal factor for the Reissner--Nordstr\"om solution describing
a charged black hole has the form
\begin{equation}                                        \label{erenom}
  N=1-\frac{2M}q+\frac{Q^2}{q^2},~~~~~~0<Q<M,
\end{equation}
where $M$ and $Q$ are the mass and charge of the black hole. The function
$N$ for a given relation between the constants can be written in the
form
\begin{equation}                                        \label{erenoq}
  N=\frac{(q-q_1)(q-q_2)}{q^2},
\end{equation}
where
$$
  q_{1,2}=M\pm\sqrt{M^2-Q^2}.
$$
Consequently, the conformal factor has double pole at $q=0$ and two
simple zeroes (horizons) at $q_{1,2}$. Its behaviour is shown in
Fig.~\ref{frenpd}.
%---------------------------------------------------------
The left branch of the conformal factor corresponds to a naked
singularity and will be not considered because the change of the order
of the pole does not affect the structure of the global solution.
Global structure of the solution for positive $q$ changes qualitatively
as compared to the Schwarzschild solution due to the existence of two
horizons. There are two conformal blocks for each of the intervals
$(0,q_1)$, $(q_1,q_2)$, and $(q_2,\infty)$. According to the rule 6)
the Carter--Penrose diagram glued from six conformal blocks and shown
in Fig.~\ref{frenpd} represents the unique fundamental region for the
Reissner--Nordstr\"om solution because the conformal factor changes its
sign inside the interval $(0,\infty)$. Its boundary consists not only
from the singular and infinite points but includes also points
corresponding to horizons. According to the rule 7) it can be periodically
continued by gluing fundamental regions along horizons in the directions
shown by arrows. In this case one obtains the universal covering space
for the Reissner--Nordstr\"om solution. The other possibility is to
identify boundary points on the lower and upper horizons after gluing
together an arbitrary number of fundamental regions.
Then the global solution will then be a cyllinder from topological
point of view.

The scalar curvature (\ref{esctmk}) for the Reissner--Nordstr\"om
solution is not constant
\begin{equation}                                        \label{escren}
  R=\frac{4M}{q^3}-\frac{6Q^2}{q^4}.
\end{equation}
%*********************************************************************
\subsection{Extremal black hole                        \label{sextbh}}
\begin{figure}[htb]%-------------------------------------------------
 \begin{center}
 \leavevmode
 \epsfxsize=120mm
 \epsfbox{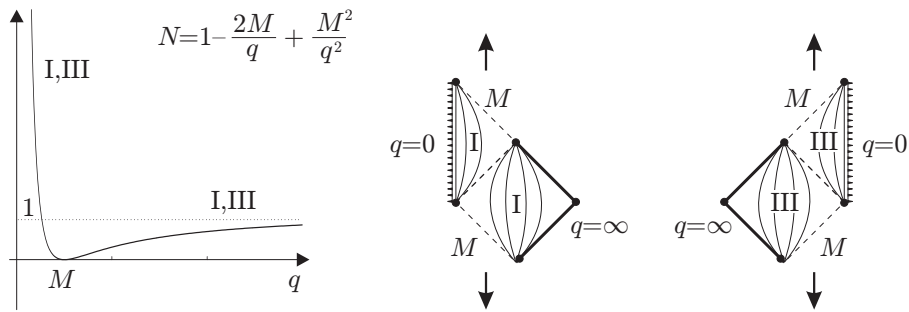}
 \end{center}
 \caption{Conformal factor and two fundamental regions for an
 extremal black hole. Arrows show directions along which solutions
 can be continued periodically.\label{fextbh}}
\end{figure}

%*********************************************************************
An extremal black hole appears from the Reissner--Nordstr\"om solution
(\ref{erenom}) when the mass equals to the charge $Q=M$. Corresponding
conformal factor
\begin{equation}                                        \label{eexblf}
  N=\frac{(q-M)^2}{q^2},
\end{equation}
has double pole at $q=0$ and double zero at $q=M$. Its behaviour is
shown in Fig.~\ref{fextbh}.
%---------------------------------------------------------
The left branch as in the previous cases describes a naked singularity.
Positive values of $q$ are devided on two intervals $(0,M)$ and $(M,\infty)$.
Two conformal blocks and their space reflection correspond to these
intervals. According to the rule 6) one has two disconnected fundamental
domains shown in Fig.~\ref{fextbh} because the conformal factor does not
change the sign. The second fundamental domain built from domains
of type {\rm III} is obtained from the domain of type {\rm I} by space
reflection. As in the case of Reissner--Nordstr\"om solution the boundary
of the fundamental domains includes horizons and can be either glued
infinitely or identified leading to the universal covering space or
cyllinders, respectively.
%*********************************************************************
\subsection{Two-dimensional gravity with torsion       \label{stwdgt}}
%*********************************************************************
The simplest geometrical gravity model in two-dimensional space-time
yeilding second order equations of motion for the zweibein and the
Lorentz connection is quadratic in curvature and torsion \cite{KatVol86}.
This model is integrable \cite{Katana89BR,Katana90}, and all solutions
are devided in two classes, the constant curvature and zero torsion
surfaces and the surfaces of nonconstant curvature and nontrivial
torsion. For nontrivial torsion the conformal factos has the form
\begin{equation}                                        \label{etwdgm}
  N(q)=q\left[(\ln^2q-2\ln q+2-\Lm)q-A\right],
\end{equation}
where $\Lm$ is the dimensionless cosmological constant and $A$ is an
arbitrary integration constant (similar to the mass in the Schwarzschild
solution). Depending on the values of these constants the conformal factor
may have up to three roots and this yeilds eleven types of global solutions
\cite{Katana93A}. For the sake of space we shall not reproduce these
solutions here. We mension that more fine classification for
two-dimensional gravity with torsion is given which takes into account
the completeness not only of extremals but geodesics too (they do not
coinside for nontrivial torsion). It takes also into account the
existence and number of degenerate extremals and geodesics.

In the present paper for simplicity we restrict ourselves to power
behaviour of the conformal factor at infinite points (\ref{ecfasi}).
The above analysis can be easily generalized to logarithmic behaviour
of the conformal factor (\ref{etwdgm}) for $q\rightarrow\infty$.
%*********************************************************************
\subsection{The dilaton gravity                        \label{sdilcf}}
\begin{figure}[htb]%-------------------------------------------------
 \begin{center}
 \leavevmode
 \epsfxsize=120mm
 \epsfbox{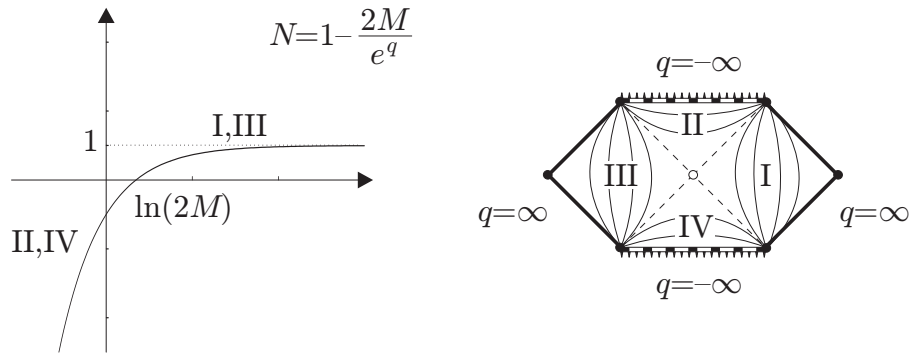}
 \end{center}
 \caption{The conformal factor and the global solution for
 two-dimensional dilaton gravity. The singularity at $q=-\infty$ is
 complete for light like extremals and incomplete for time like ones.
 \label{fdilat}}
\end{figure}

%*********************************************************************
Two-dimensional dilaton gravity model is described by the metric and
the scalar dilaton field. It is closely related to string theory and is
integrable \cite{Witten91,MaSeWa91}.
The conformal factor in this case has the form
\begin{equation}                                        \label{ecofdi}
  N(q)=1-\frac{2M}{e^q}.
\end{equation}
For positive $M$ the conformal factor has one horizon, and the global
solution is similar to the Schwarzschild black hole,
see.\ Fig.~\ref{fdilat}.
%---------------------------------------------------------
The exponential behaviour of the conformal factor (\ref{ecofdi}) near the
curvature singularity when $q\rightarrow-\infty$ leads to qualitative
distinction of this solution from the black hole because the singularity
is semi complete. Namely, all light like extremals approach the singularity
at an infinite value of the canonical parameter while time and space
like extremals meet the singularity at a finite value \cite{KaKuLi97}.

Note that the dilaton gravity is locally equivalent to quadratic
two-dimensional gravity with torsion but global equivalence is
abcent \cite{KaKuLi96}.
%*********************************************************************
\subsection{Minkowskian plain                          \label{sminpl}}
%*********************************************************************
The above examples demonstrate the rules of construction of global
Lorentz surfaces of nonconstant curvature. The following two examples
show how the rules work for constant curvature surfaces, the classical
field of differential geometry. Example of the Minkowskian plain is
also important because it will be used later in the proof of the
smoothness of global solutions in saddle points.

For the Minkowskian plain the scalar curvature is equal to zero, and
the conformal factor is a linear function
\begin{equation}                                        \label{ecfmis}
  N=bq+c,~~~~~~b,c=\const.
\end{equation}
There are two qualitatively different cases $b=0$ and $b\ne0$.

\begin{figure}[htb]%-------------------------------------------------
 \begin{center}
 \leavevmode
 \epsfysize=32mm
 \epsfbox{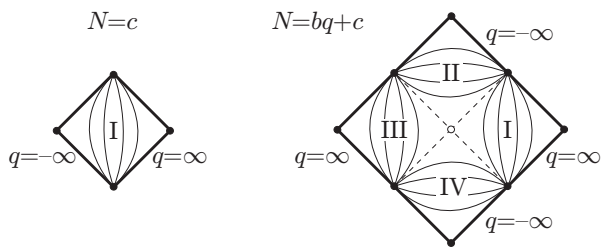}
 \end{center}
 \caption{Two representations of the Minkowskian plain in the absence
 and presence of one horizon. In the last case the Minkowskian plain
 is sewed from four conformal blocks.\label{fminpl}}
\end{figure}

For $b=0$ the metric (\ref{emetok}) is the Minkowskian one
\begin{equation}                                        \label{eminmo}
  ds^2=c(d\tau^2-d\s^2),
\end{equation}
where we assume $c>0$ for definiteness. The conformal factor
$N=c$ does not have nor singularities nor zeroes. Then equation
(\ref{eshiff}) yeilds
$$
  q=\pm c\s.
$$
Here the $\pm$ sign corresponds to different orientation of the variable
$q$ with respect to space coordinate. Since this variable does not enter
the metric, it can be excluded from the consideration. Hence one
square conformal block shown in Fig.~\ref{fminpl} corresponds to the
interval $q\in(-\infty,\infty)$.
%---------------------------------------------------------

Let us consider nonzero $b>0$. Then the conformal factor (\ref{ecfmis})
has one simple zero at $q_1=-c/b$. Two pairs of conformal blocks
correspond to the intervals $(-\infty,q_1)$ and $(q_1,\infty)$ wich are
sewed together along horizons as shown in Fig.~\ref{fminpl}. Let us show
how this solution is related to the previous representation of the
Minkowskian plain by one conformal block. Without loss of generality we
set $c=0$, that is
\begin{equation}                                        \label{ecomph}
  ds^2=b|q|(d\tau^2-d\s^2)
\end{equation}
This can be always achieved by the shift $q\rightarrow q+\const$, which
does not affect the global structure of the solution. Consider all four
domains

{\bf Domain {\rm I}.} For $N=bq$ equation (\ref{eshiff}) yeilds
\begin{equation}                                        \label{eqsfir}
  q=e^{b\s}
\end{equation}
up to a shift of $\s$. Corresponding metric is static, has the form
\begin{equation}                                        \label{emifir}
  ds^2=be^{b\s}(d\tau^2-d\s^2),~~~~~~\tau,\s\in\MR^2,
\end{equation}
and is defined on the whole plain $\tau,\s\in(-\infty,\infty)$. Transition
to the light cone coordinates (\ref{elicco}) and the conformal
transformation
\begin{equation}                                        \label{ectrfi}
  u=\frac2be^{\frac{b\x}2}>0,~~~~~~
  v=-\frac2be^{-\frac{b\eta}2}<0
\end{equation}
yield the Minkowskian metric
\begin{equation}                                        \label{emimec}
  ds^2=b\,dudv,
\end{equation}
defined on the first quadrant.

{\bf Domain {\rm III}}. For the variable $q$ and the metric one has
\begin{eqnarray}                                        \label{eqsthi}
  q&=&e^{-b\s},
\\                                                      \label{emithi}
  ds^2&=&be^{-b\s}(d\tau^2-d\s^2),~~~~~~\tau,\s\in\MR^2.
\end{eqnarray}
The conformal transformation leading to the Minkowskian metric
(\ref{emimec}) has the form
\begin{equation}                                        \label{ectrth}
  u=-\frac2be^{-\frac{b\x}2}<0,~~~~~~
  v=\frac2be^{\frac{b\eta}2}>0,
\end{equation}
corresponding to the third quadrant.

{\bf Domain {\rm II}}. This domain is homogeneous
\begin{eqnarray}                                        \label{eqssec}
  q&=&-e^{b\tau},
\\                                                      \label{emisec}
  ds^2&=&be^{b\tau}(d\tau^2-d\s^2),~~~~~~\tau,\s\in\MR^2.
\end{eqnarray}
The conformal transformation yielding the Minkowskian metric has the
form
\begin{equation}                                        \label{ectrse}
  u=\frac2be^{\frac{b\x}2}>0,~~~~~~
  v=\frac2be^{\frac{b\eta}2}>0,
\end{equation}
corresponding to the second quadrant.

{\bf Domain {\rm IV}}. Analogously, for the variable $q$ and the metric
one has
\begin{eqnarray}                                        \label{eqsfor}
  q&=&-e^{-b\tau},
\\                                                      \label{emifor}
  ds^2&=&be^{-b\tau}(d\tau^2-d\s^2),~~~~~~\tau,\s\in\MR^2.
\end{eqnarray}
The conformal transformation leading to the Minkowskian metric is
\begin{equation}                                        \label{ectfor}
  u=-\frac2be^{-\frac{b\x}2}<0,~~~~~~
  v=-\frac2be^{-\frac{b\eta}2}<0,
\end{equation}
corresponding to the fourth quadrant.

For all four domains the horizon $q_1=0$ corresponds to the coordinate
axes $u=0$ and $v=0$. Simple calculations show that in all domains
\begin{equation}                                        \label{emisqc}
  q=-\frac{b^2}4(t^2-x^2),
\end{equation}
where
$$
  u=t+x,~~~~~~v=t-x.
$$
These coordinates yeild the simplest example of the Kruskal--Szekeres
coordinates for the metric (\ref{ecomph}). That is the variable $q$
is the hyperbolic radius of the polar coordinate system on the
Minkowskian plain. For the hyperbolic polar angle in the static
domains
$$
  \f=\arcth\left(\frac tx\right),
$$
one has
$$
  {\rm I}:~~\f=\frac b2\tau,~~~~~~{\rm III}:~~\f=-\frac b2\tau.
$$
In the homogeneous domains
$$
  \f=\arcth\left(-\frac xt\right),
$$
and
$$
  {\rm II}:~~\f=-\frac b2\s,~~~~~~{\rm IV}:~~\f=\frac b2\s.
$$

Hence conformal transformations (\ref{ectrfi}), (\ref{ectrth}),
(\ref{ectrse}), and (\ref{ectfor}) yield one and the same Minkowskian
metric (\ref{emimec}) in all four domains but defined on different
quadrants, the smoothness of the metric on the horizons and the saddle
point for the $q$ variable (\ref{emisqc}) is proved.

Thus the linear conformal factor (\ref{ecfmis}) for $b\ne0$
describes the Minkowskian plain, Fig.\ \ref{fminpl}, {\it b}, in the
hyperbolic coordinate system.
%*********************************************************************
\subsection{Constant curvature surfaces                \label{scocus}}
%*********************************************************************
Let us consider constant curvature surfaces for which the conformal
factor is a quadratic polinomial (\ref{ecocsp}). For definiteness we
consider positive curvature surfaces $a>0$. Negative curvature surfaces
are obtained from the positive curvature ones by the transformation
$\tau\leftrightarrow\s$, that is by the rotation all the diagrams by
the angle $\pi/2$. Depending on the values of the constants $a,b,c$
equation $N(q)=0$ may not have, have one or two zeroes. Consider all
three cases leading to different metrics and Carter--Penrose diagrams
for the constant curvature surfaces consequtively. Without loss of generality
we set $b=0$, which can be always achieved by the shift of $q$.

\begin{figure}[htb]%-------------------------------------------------
 \begin{center}
 \leavevmode
 \epsfxsize=120mm
 \epsfbox{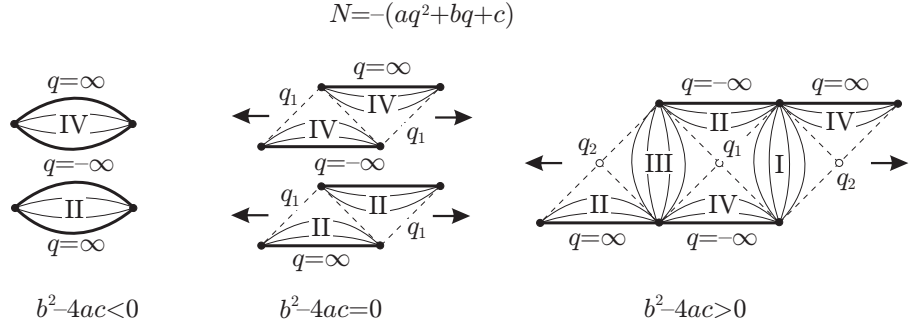}
 \end{center}
 \caption{Carter--Penrose diagrams for constant positive curvature
 surfaces. There are three cases: the absence of a horizon, the
 existence of one or two horizons.
 \label{fcocsu}}
\end{figure}

{\bf Absence of a horizon.}
\begin{equation}                                        \label{ecocsw}
  N=-(aq^2+c),~~~~~~a>0,~c>0.
\end{equation}
Since $N<0$ and singularities and zeroes are absent, the global solution
corresponds to an infinite interval $q\in(-\infty,\infty)$. It is
represented by the homogeneous conformal block of type {\rm II} or
{\rm IV} shown in Fig.~\ref{fcocsu}.
%---------------------------------------------------------
In the domain {\rm IV} equation (\ref{eshiff}) for the conformal factor
(\ref{ecocsw}) yeilds
\begin{equation}                                        \label{ecjcso}
  q=\sqrt{\frac ca}\,\tg(\sqrt{ac}\,\tau)
\end{equation}
up to a shift of $\tau$. Time coordinate varies in the finite interval
$$
  -\frac\pi2<\sqrt{ac}\,\tau<\frac\pi2,
$$
corresponding to a lens type diagram. The metric in the $\tau,\s$
plain has the form
\begin{equation}                                        \label{eccmef}
  ds^2=\frac c{\cos^2(\sqrt{ac}\,\tau)}(d\tau^2-d\s^2).
\end{equation}
This metric is a global one and defined on the whole universal covering
space of constant curvature. In the domain {\rm II} the metric has
the same form except the orientation of $q$ with respect to $\tau$
changes.

{\bf One horizon.}
\begin{equation}                                        \label{ecohsw}
  N=-aq^2,~~~~~~a>0.
\end{equation}
At the point $q_1=0$ the conformal factor has a zero of second order
corresponding to a horizon. In domain {\rm IV} for both intervals
$q\in(-\infty,0)$ and $q\in(0,\infty)$ one has
\begin{eqnarray}                                        \label{eqfccs}
  q&=&-\frac1{a\tau},
\\                                                      \label{eincct}
  ds^2&=&\frac1{a\tau^2}(d\tau^2-d\s^2).
\end{eqnarray}
In this case the fundamenral region consists of two triangular conformal
blocks shown in Fig.~\ref{fcocsu},{\it b}. To get the universal covering
space diffeomorphic to the diagram of lens type from the previous case
the fundamental domain should be extended in both directions. Identifying
left and right boundary of the fundamental domain one gets the surface
diffeomorphic to the one sheet hyperboloid (see, for example,
\cite{Katana93A}).

{\bf Two horizons.}
\begin{equation}                                        \label{ecthsw}
  N=-(aq^2+c),~~~~~~a>0,~c<0.
\end{equation}
This conformal factor has two simple zeroes at the points
$q_{1,2}=\mp\sqrt{-c/a}$. Hence there are six conformal blocks for
three intervals. The symmetry of the conformal factor imply that
four homogeneous and two static conformal blocks are essentially
the same. For a homogeneous domain {\rm IV} one has
\begin{eqnarray}                                        \label{eqthcs}
  q&=&-\sqrt{-\frac ca}\,\cth(\sqrt{-ac}\,\tau),
\\                                                      \label{einfth}
  ds^2&=&\frac{-c}{\sh^2(\sqrt{-ac}\,\tau)}(d\tau^2-d\s^2).
\end{eqnarray}
For static domain {\rm I} one has
\begin{eqnarray}                                        \label{eqshcs}
  q&=&\sqrt{-\frac ca}\,\tanh(\sqrt{-ac}\,\s),
\\                                                      \label{eisfth}
  ds^2&=&\frac{-c}{\ch^2(\sqrt{-ac}\,\s)}(d\tau^2-d\s^2).
\end{eqnarray}

Different forms of the metric on a constant curvature surface are caused
by different coordinate choices. Since all metrics are conformally flat
they are related by conformal coordinate transformations. Let us write
metric (\ref{eincct}) in light cone coordinates (\ref{elicco}) and make
a conformal transformation $\x\rightarrow u(\x)$,
$\eta\rightarrow v(\eta)$, then
\begin{equation}                                        \label{ecotrm}
  ds^2=\frac1{a\tau^2}(d\tau^2-d\s^2)
  =\frac4{a(\x+\eta)^2}\frac{d\x}{du}\frac{d\eta}{dv}dudv.
\end{equation}
The transformation
\begin{equation}                                        \label{efcotr}
  \x=\frac2{\sqrt{ac}}\tg\left(\frac{\sqrt{ac}}2u\right),~~~~~~
  \eta=-\frac2{\sqrt{ac}}\ctg\left(\frac{\sqrt{ac}}2v\right)
\end{equation}
yields the metric (\ref{eincct}). Transformations
\begin{equation}                                        \label{esectx}
  \x=\frac2{\sqrt{-ac}}\tanh\left(\frac{\sqrt{-ac}}2u\right),~~~~~~
  \eta=\frac2{\sqrt{-ac}}\tanh\left(\frac{\sqrt{-ac}}2v\right)
\end{equation}
and
\begin{equation}                                        \label{etectx}
  \x=\frac2{\sqrt{-ac}}\tanh\left(\frac{\sqrt{-ac}}2u\right),~~~~~~
  \eta=-\frac2{\sqrt{-ac}}\cth\left(\frac{\sqrt{-ac}}2v\right)
\end{equation}
yeild metrics (\ref{einfth}) and (\ref{eisfth}), respectively.
%*********************************************************************
\section{Eddington--Finkelstein coordinates            \label{sedfic}}
%*********************************************************************
The smoothness of the metric on horizons is proved by writing it in a
new coordinate system covering horizons. Let the zero $q_j$ of an odd
order be inside the interval $(q_-,q_+)$. For definiteness we assume that
the conformal factor is positive and negative in the adjacent intervals
$(q_{j-1},q_j)$ and $(q_j,q_{j+1})$, respectively. In accordance with the
rule 4) the stationary conformal block {\rm I} must be sewed together
with two homogeneous blocks of types {\rm II} and {\rm IV}. Horizons can
be covered by the {\em Eddington--Finkelstein coordinates}
\cite{Edding24A,Finkel58},
\index{Eddington--Finkelstein coordinates}%
introduced in the following way.

{\bf Sewing together domains {\rm I}--{\rm II}.} To prove the smoothness
of the sewing together domains {\rm I} and {\rm II} by themselves and the
metric on them the Eddington--Finkelstein coordinates will be introduced.
These coordinates cover the chosen domains, the transition functions in each
of the domain being in the ${\cal C}^{l+1}$ class, and the metric in new
coordinates including horizons is in ${\cal C}^l$ class.

Let us go from the conformal coordinates $\tau,\s$ on the domain {\rm I}
to the Eddington--Finkelstein coordinates $\x,q$
\begin{equation}                                        \label{eedfit}
  \tau = \x-{\displaystyle\int^q\frac{dr}{N(r)}},~~~~~~
  \s   = {\displaystyle\int^q\frac{dr}{N(r)}}.
\end{equation}
The last integral is divergent on horizons but this is not important
because the transformation (\ref{eedfit}) is considered only in the
interior points of the domain {\rm I}. The transition functions are
obviously of ${\cal C}^{l+1}$ class in the interior points.
Substituting expressions for the differentials
\begin{equation}                                        \label{edfidt}
  d\tau = d\x-\displaystyle{\frac{dq}{N}},~~~~~~
  d\s   = \displaystyle{\frac{dq}{N}},
\end{equation}
in the metric (\ref{emetfd}) one gets the quadratic form
\begin{equation}                                        \label{edfint}
  ds^2=Nd\x^2-2dqd\x.
\end{equation}
Determinant of this metric is equal to $\det g_{\al\bt}=-1$, and the
metric (\ref{edfint}) is defined for all $\x\in(-\infty,\infty)$ and,
importantly, for all $q\in(q_-,q_+)$ and not only for $q\in(q_{j-1},q_j)$.
That is the metric (\ref{edfint}) is defined in a larger domain then
the starting one (\ref{emetok}). The smoothness of the metric (\ref{edfint})
coinsides with the smothness of the conformal factor everywhere including
horizons.

In the homogeneous domain of the type
\begin{equation}                                        \label{emetht}
  \text{II}:~~ds^2=-N(q)(d\tau^2-d\s^2),~~~~\frac{dq}{d\tau}=N(q)<0,
\end{equation}
we go to the Eddington--Finkelstein coordinates
\begin{equation}                                        \label{edfict}
  \tau = {\displaystyle\int^q\frac{dr}{N}},~~~~~~
  \s   = \x-{\displaystyle\int^q\frac{dr}{N}}.
\end{equation}
The coordinates transformation in the domain {\rm II} differs from
the transformation (\ref{eedfit}). However the metric (\ref{emetht})
takes the form (\ref{edfint}). This means that Eddington--Finkelstein
coordinates with a given metric (\ref{edfint}) isometrically cover
domains {\rm I} and {\rm II}. This proves that the smoothness of the
metric on the horizon coinsides with the smoothness of the conformal
factor. One may check that the horizon $q_j$ is an extremal by itself
in the Eddington--Finkelstein coordinates.

The domains {\rm III}--{\rm IV} can be sewed analogously. Explicit
formulas for the transition to Eddington--Finkelstein coordinates have
the form
\begin{equation}                                        \label{eedfth}
{\rm III}:~~~~~~
  \tau = \x+{\displaystyle\int^q\frac{dr}{N(r)}},~~~~~~
  \s   = -{\displaystyle\int^q\frac{dr}{N(r)}}.
\end{equation}
and
\begin{equation}                                        \label{edfifo}
{\rm IV}:~~~~~~
  \tau = -{\displaystyle\int^q\frac{dr}{N}},~~~~~~
  \s   = \x+{\displaystyle\int^q\frac{dr}{N}}.
\end{equation}
The metric has the same form in both domains
\begin{equation}                                        \label{edfitf}
  ds^2=Nd\x^2+2dqd\x.
\end{equation}
Metrics (\ref{edfint}) and (\ref{edfitf}) are connected between
themselves by the transformation $\x\rightarrow-\x$.

{\bf Sewing together domains {\rm I}--{\rm IV}.}
In the domain {\rm I} we choose the coordinates
\begin{equation}                                        \label{edfiff}
  \tau = \eta+{\displaystyle\int^q\frac{dr}{N(r)}},~~~~~~
  \s   = {\displaystyle\int^q\frac{dr}{N(r)}},
\end{equation}
in which the metric takes the form
\begin{equation}                                        \label{eintff}
  ds^2= Nd\eta^2+2dqd\eta.
\end{equation}
The coordinates
\begin{equation}                                        \label{edffou}
  \tau = -{\displaystyle\int^q\frac{dr}{N(r)}},~~~~~~
  \s   = -\eta-{\displaystyle\int^q\frac{dr}{N(r)}}
\end{equation}
yield the same metric in the domain {\rm IV}.

Similar coordinates are introduced for sewing together the domains
{\rm II} and {\rm III}, the corresponding metric being related to
the metric (\ref{eintff}) by the transformation $\eta\rightarrow-\eta$.

If the horizon $q_j$ has even degree and in both intervals $(q_{j-1},q_j)$
and $(q_j,q_{j+1})$ and the conformal factor is, for example, positive, then
according to the rule 4) the blocks of one type are only sewed together.
To prove the smoothness of the sewing on the blocks of type {\rm I} and
{\rm III} the coordinates (\ref{eedfit}) and (\ref{eedfth}) are introduced
respectively.

Thus the smoothness of the metric and, consequently, Christoffel's symbols
and curvature tensor is proved on all horizons. Metric in the
Eddington--Finkelstein coordinates (\ref{edfint}) covers the whole chain
of the conformal blocks for $q\in(q_-,q_+)$ of types {\rm I} and {\rm II}
for the fundamental domain. For this chain the coordinate $q$ decreases
with the increasong of $\tau$. The chain is a $\CC^{l+1}$ manifold
because the transition functions (\ref{eedfit}), (\ref{edfict}) are of
this class. Metric (\ref{edfitf}) covers parallel chain of the conformal
blocks {\rm III} and {\rm IV}. Metric in the form (\ref{eintff}) covers
a perpendicular chain of the conformal blocks {\rm I} and {\rm IV}.
In the interior of every chain the metric in the Eddington--Finkelstein
coordinates is nondegenerate and of the ${\cal C}^l$ class. The fundamental
domain consists of two parallel chains of conformal blaocks. If there is
at least one zero of odd degree inside the interval, then the fundamental
domain is connected. In the opposite case one has two disconnected
fundamental domains. This is the meaning of the rule 5).

The Eddington--Finkelstein coordinates are natural ones in the following
sence. First of all note that the extension of the solution has to
be performed along variable $q$ because it defines the completeness
of the manifold and the curvature singularities, and it is a scalar
function in addition. Since for a static solution the variable
$q$ depends only on $\s$ the $q$ variable (\ref{eedfit}), (\ref{eedfth}),
(\ref{edfiff}), may be chosen as a coordinate instead of $\s$, the
coordinates being one to one related by the equation (\ref{eshiff}) up
to an insignificant shift of $\s$ in each domain. After that the light
cone coordinate (\ref{elicco}) is introduced instead of the time
coordinate. There are two possibilities, and both of them are realised
yeilding coordinates on the two perpendicular chains of conformal blocks.
The Eddington--Finkelstein coordinates on the homogeneous blocks are
introduced analogously. One may check that there is no possibility
to sew together the domains {\rm I}, {\rm III} or {\rm II}, {\rm IV}.
Moreover straightforward calculations prove that the derivatives of
$q$ are discontinuos under this sewing while $q$ by itself is a
continuous function \cite{Katana93A}.
%*********************************************************************
\section{Smoothness of the metric in the saddle point  \label{sedpom}}
%*********************************************************************
The Eddington--Finkelstein coordinates do not cover the saddle points
been situated on the horizons crossing. Saddle points correspond to
zeroes of $N(q)$ situated at finite points with the exponent $m=1$ or
$m\ge2$ (see Fig.~\ref{fboupr}). If the point $q_j$ is a zero
of second order or higher then the saddle point is complete and
nothing has to be proved. If the point $q_j$ is a simple zero then one may
introduce coordinates covering a neiborhood of the saddle point.
Since the behaviour of the conformal factor near the zero is the same
as for the Minkowskian plain we go to the Kruskal--Szekeres coordinates
with $b=N'(q_j)$ (see section \ref{sminpl}). Explicit transformation
formulas in domains {\rm I}--{\rm IV} have the form (\ref{ectrfi}),
(\ref{ectrse}), (\ref{ectrth}), and (\ref{ectfor}), respectively.
As for the Minkowskian plain the $u,v$ coordinates cover a neighborhood
of the saddle point including the domains of all four types sewed together
along horizons. In all four quadrants metric (\ref{emetok}) takes the
same form
\begin{equation}                                        \label{emesap}
  ds^2=|N|\exp\left(-\int^q dp\frac{N'(q_j)}{N(p)}\right)\,dudv
  =\exp\left(\int^q dp\frac{N'(p)-N'(q_j)}{N(p)}\right)\,dudv.
\end{equation}
Since the limit of the
integrand exists in the saddle point the integral is a smooth function.
This proves nondegeneracy and smoothness of the metric in the saddle
point, the differentiability of the metric being the same as of the
conformal factor. The Kruskal--Szekeres coordinates can be obviously
introduced near every saddle point of first order. Because the
transition functions to the Kruskal--Szekeres coordinates are of
$\CC^\infty$ class, then the $\CC^{l+1}$ smoothness class of the
universal covering space is defined by the transition functions to
the Eddington--Finkelstein coordinates. This finishes the proof
of the main theorem \ref{tunisp}.
%*********************************************************************
\section{Conclusion                                    \label{sconcl}}
%*********************************************************************
The method of building of global solutions for two-dimensional
metrics of Lorentz signature considered in the present paper is
simple and constructive. If the gravity equations of motion lead to
the metric (\ref{emetok}) with some conformal factor then it is sufficient
to analyse its zeroes and singularities. Afterwards the universal covering
space is uniquely constructed. All other global solutions are factor
spaces of the universal covering space under the action of free and
descrete transformation group. The possibilities to construct solutions
with nontrivial fundamental groups by the identification of the boundary
of the fundamental domain are noted in the paper. In a general case
the finding of descrete transformation groups depends on the form of the
conformal factor and is the subject of independent investigation.

The author is very gratefull to T.~Kl\"osch, W.~Kummer, T.~Strobl,
I.~V.~Volovich, V.~V.~Zharinov for fruitfull discussions on the subject
and the Russian Fund for basic research, grants RFBR-99-01-00866 and
RFBR-96-15-96131 for financial support.

%\bibliography{my,book,2dgrav,3dgrav,defect,math,string,kalkle,gravity}
%\bibliographystyle {unsrt}
\end{document}